\documentclass[runningheads]{llncs}
\usepackage{graphicx}
\begin{document}
\title{Retracing the Flow of the Stream: Investigating Kodi Streaming Services}


\author{Samuel Todd Bromley\inst{1} \and
John Sheppard\inst{2,3} \and
Mark Scanlon\inst{3} \and
Nhien-An Le-Khac\inst{3}}

\authorrunning{Bromley et al.}
\institute{Royal Canadian Mounted Police, Canada
\email{stbromley@gmail.com}\\\and
Waterford Institute of Technology, Ireland
\email{jsheppard@wit.ie}\\ \and
Forensics and Security Research Group, University College Dublin, Ireland\\
\email{\{mark.scanlon, an.lekhac\}@ucd.ie}}

%
%
%
%
%
\maketitle              
\begin{abstract}
Kodi is of one of the world's largest open-source streaming platforms for viewing video content.
Easily installed Kodi add-ons facilitate access to online pirated videos and streaming content by facilitating the user to search and view copyrighted videos with a basic level of technical knowledge. In some countries, there have been paid child sexual abuse organizations publishing/streaming child abuse material to an international paying clientele. Open source software used for viewing videos from the Internet, such as Kodi, is being exploited by criminals to conduct their activities.
In this paper, we describe a new method to quickly locate Kodi artifacts and gather information for a successful prosecution. We also evaluate our approach on different platforms; Windows, Android and Linux. Our experiments show the file location, artifacts and a history of viewed content including their locations from the Internet. Our approach will serve as a resource to forensic investigators to examine Kodi or similar streaming platforms.

\end{abstract}
%
%
%

\section{Introduction}
\label{intro}


The rise in illegal streaming services saw the September 2019 shutdown of the Xtream-codes IPTV service. This service was run from Greece and operated through servers in Italy, the Netherlands, France, Germany and Bulgaria. It has been reported that the operation involved over two hundred servers and one hundred and fifty PayPal accounts used by the criminals. It is estimated that there were 5,000 re-sellers of the service to over 50 million subscribers~\cite{Bouma:2019}. 

One of the most popular streaming applications is Kodi. Kodi is an open source media center developed by non-profit technology consortium XBMC. It is compatible with Windows, Linux, OSX and Android. It is used for streaming music and video from a user's private library on their own network or from public sources over the Internet. It is highly customizable through the use of add-ons. Add-ons, written by third-party developers, are easily installed through the Kodi interface by specifying a repository or ``repo'' URL to connect to. 

In this paper, we aim to answer the following research questions: 1) How should a forensic investigation of of Kodi be conducted? 2) What can be recovered from a Kodi system? 3) Where are the Kodi forensic artifacts located? The contribution of this work can be summarized as follows:
\begin{itemize}
    \item Introducing a framework for conducting IPTV investigations based on Kodi.
    \item Detailing the Kodi ecosystem where evidence can be found for investigation of digital piracy or streaming of illegal content.

\end{itemize}




\section{Related Work}
\label{related}


\subsubsection{IPTV Forensics -}
Users purchase IPTV subscriptions from commercial sellers, which grants access to the sellers' private server's live television feeds using IPTV set top boxes or software. Alternatively, users can download software to access less reliable free streams. The cybercrime aspect of IPTV and cloud investigations are examined in~\cite{Sheppard:2018}. A case study was conducted, focusing on the cloud based development and distribution networks of the most popular Kodi Add-ons of 2018. Acestreams are an example IPTV service using streaming mesh networks~\cite{Hei:2008}. Acestreams employ the use of BitTorrent technology to distribute live television, the principle being, that one user's upstream data pipe, is used to supply video content to another user's downstream pipe to view video~\cite{scanlon2010week}. 


\subsubsection{Android Set Top Box Forensics -}

Android set top boxes are by far the most popular method of using Kodi. These low cost devices are sold online and are pre-configured to run Kodi and other streaming software. Android-based forensic investigations have primarily focused on mobile phone and tablet versions of Android and applications installed on these devices~\cite{Sgaras:2015,Walnycky:2015}. An overview of Android forensics and the Android file system can be found in~\cite{Hoog:2011} and~\cite{Faheem:2014}. These provide a framework for Kodi artifacts on the Android operating system. Given the growth of Android powered Smart TVs, Kodi can be expected to become more common place.

\section{Findings and Analysis for Kodi on Windows}

\subsubsection{Media Library -}
The Kodi media library is a feature of Kodi that allows the cataloging and sharing of downloaded movies and videos that are located on the computer. This feature is used for personal video collections. The media library tracks which videos have been previously played, allowing the user to watch videos without repetition. This is an excellent source of information for investigators. All previously played videos are stored in the media library artifacts, including those from external media. This log of previously viewed media could inform the investigator of hidden portable media that may not have been seized.


\subsubsection{Add-ons -}
The true benefits of Kodi are the user created add-ons. User created Add-ons allow the user to view copyrighted material hosted for free on the Internet. These add-ons allow the Kodi user to easily navigate, search and stream copyrighted materials. The MPAA and the ACE sub group are actively attempting to have these add-ons eliminated. As soon as one is taken down legally another typically pops up.




\subsubsection{Artifacts -}


The minimum number of pages detected within the databases is eight when no files have been indexed. The file metadata records have all been observed to start from page 8 of the database on-wards. In general, data within the pages are stored little-endian (least significant byte first), however big-endian values have been noted as existing within individual records.


The SQLite database ``Addons27.db'' is located in \texttt{C:\textbackslash Users\textbackslash X\textbackslash AppData\textbackslash} \texttt{ Roaming\textbackslash Kodi\textbackslash userdata\textbackslash Database\textbackslash Addons27.db}. This table contains artifacts related to add-ons including associated repositories. These add-on artifacts identify the name of installed add-ons, whether the add-on is still in use, installation date of the installed add-ons, date add-on updated, date add-on last used, and the origin of the add-on.



\subsubsection{Repositories -}

Repository.exodusredux is the Repository that was manually installed in .zip form. ``b6a50484-93a0-4afb-a01c-8d17e059feda'' is the serial number of the original installer file. This indicates that the file was an original, non-updated, file loaded from the Kodi installer~\cite{Addons:Doc}. In the ``Addons27.db'' SQLite database there is a table called ``repo''. The `repo'' table provides the following artifacts; name of repository installed, checksum for installed repository, last update check date, and the version of the repository.



\subsubsection{Kodi Log Files -}
Kodi activity is recorded in two log files. One is a current log file active since the latest reboot, the other is an inactive log file from the previous reboot. In the Windows environment, Kodi saves a log file in the following location: \texttt{C:\textbackslash Users\textbackslash X\textbackslash AppData\textbackslash} \texttt{Roaming\textbackslash Kodi\textbackslash kodi.txt}. This file contains all actions by the user for that session. From a forensic point of view, interesting artifacts include: date and time Kodi last used, searches performed, user account, last viewed file, source of last viewed file, and when was Kodi last shut down. As Kodi overwrites its oldest log file on starting the application information will be lost if the investigator executes the Kodi program in a non-forensically sound manner.

\section{Examining Kodi installed on Android and Linux}

\subsubsection{Android -}
Our experiments were conducted using a Samsung S9+ Android mobile device running the Android 9.0 OS, as the test Android system to locate the Android artifacts. The artifacts and databases are located in similar locations in the Android operating system as they are in other operating systems.

In the Kodi userData folder (hidden by default), all the same database Kodi artifacts are present which can be logically extracted.


\subsubsection{Linux (Ubuntu) -} The Kodi file structure once again is the same. The default Kodi installation with that version of Ubuntu is Kodi version 15.2 (Isengard) in the hidden folder ``.kodi''. The folder and databases structure is identical with the minor difference that in this version of Kodi, the artifacts are called ``MyVideos93.db'' and ``Addons19.db''. The 2 numerical digits at the end of the database name are the result of the Kodi version. The information contained is identical. The database is stored ``.kodi'' folder. Although the SQLite database names are slightly different, the contents are the same.

\subsubsection{Linux (OSMC) -}

The last device investigated was a Raspberry Pi 3 running Open Source Media Center (OSMC). OSMC is a dedicated Kodi Linux operating system. The SD card used was first sanitized and OSMC\_TGT\_rbp2\_20180316 was installed through OSMC's qt\_host\_installer application. The Pi was configured for a wired network and debugging was enabled on the device. A selection video add-ons were again installed and experiments were conducted as before. Once this was completed, the SD card was removed from the Raspberry Pi and a raw image was created using the command-line tool, $dd$. The image was analyzed using Autopsy and FTK Imager for dual tool verification. The purpose of the analysis was to investigate the filesystem layout, the structure of the add-ons, the location of Kodi logs and the identification of Kodi artifacts.

\section{Discussion}

\subsubsection{Kodi Home Folder -}
The Kodi directory is the most important directory of a Kodi investigation. This location contains the following directories:

\begin{itemize}
    \item add-ons - a list of repository add-ons and video add-ons associated with the device under examination
    \item media - media files
    \item system - system customization, empty by default
    \item temp - log files
    \item userdata - databases, settings files and customization files
\end{itemize}

The add-ons folder contains a set of sub folders classed as repositories, video plugins, movie and TV metadata scrapers, language resources, and scripts. It also contains a packages folder that contains zipped versions of these files with timestamps of when they were downloaded. 

\subsubsection{Kodi Add-ons -}

Investigation of these add-ons reveals a common core file structure comprising of .xml, python, .txt and image files. These add-on directories typically contain the following files and reveal the following information:


\begin{itemize}
  \item The file \emph{add-on.xml} comprises of several metadata fields that describe the add-on to the user and system. Other fields reference any required dependencies, credits and version information. It also includes the update URL.
  \item The file \emph{add-on.py} is the Python code for that add-on.
  \item The \emph{resources} sub directory stores add-on files that do not need to be stored in the root directory, such as software libraries, translations, and images.
  \item The file \emph{License.txt} contains the text of the add-on's software license~\cite{Kodi:2019}.
\end{itemize}



\subsubsection{Kodi Userdata -}

The userdata folder contains the following folders: 

\begin{itemize}
    \item add-on data - contains configuration data for the add-ons installed in the .kodi/add-ons directory
    \item database - containing databases that are required for locating Music and Video Libraries and any downloaded or scraped music or video information
    \item keymaps - contains files for any customized keymapping
    \item library - custom libraries
    \item peripheral data
    \item playlists - where the playlists are stored
    \item thumbnails - cached thumbnails
\end{itemize}

\subsubsection{Kodi Databases -}

The database files can be found in the userdata/Database location. The following databases are present in SQLite format. Different versions of Kodi use different numbers appended to the end of the SQLite databases. In each instance the number $XX$ is determined by the current Kodi version in use.

\begin{itemize}
    \item Addons$XX$.db - Database file containing information on all Kodi add-ons
    \item ADSP$XX$.db-Database file containing information for Audio Digital Signal Processing add-ons
    \item EPG$XX$.db - Database file containing information on Electronic Programme Guide (EPG) for Live-TV
    \item MyMusic$XX$.db - Database file containing Music information
    \item MyVideos$XX$.db - Database file containing Movie, TV Show and Music Videos information
    \item Textures13.db - Database file containing information on all Kodi thumbnails, fanart and posters.
    \item TV$XX$.db - Database file containing information on Live-TV channels
    \item ViewModes$XX$.db - Database file containing information on the use and frequency of the Kodi device.
\end{itemize}


\subsubsection{Kodi Logs -} 

In the event of a Kodi machine being investigated, care should be taken around the Kodi logs. If the machine is switched off, the machine should first be imaged using appropriate controls. This will prevent the log from before the last reboot from being overwritten. This will inform an investigator with information such as when an add-on was installed on the device or recent activity that has occurred on the device.







\section{Conclusion and Future Work}
\label{conclusion}
 
 
 
This paper has compared the Kodi filesystem, logs, add-ons, databases and other artifacts across a range of devices. It examined Kodi as a software application on a Windows 10 and Ubuntu machine, the Kodi Android app, Kodi as a stand-alone operating system, and OSMC. The primary sources where evidence could be collected were highlighted through the Kodi folder. The structure of the Kodi filesystem has been presented. Future work in this area requires the investigation of the relationships with these add-ons and the cyberlockers providing the illegal streams. In addition, much remains to be done with automated stream content analysis using computer vision to automatically detect illegal content~\cite{10.1145/3407023.3407068,anda2020UnderageAgeEstimation}. As techniques are created to monetize such illicit services by cybercriminals, the need for the investigation of these machines will increase.

\bibliographystyle{splncs04}
\bibliography{bibfile.bib}
%




\end{document}